\documentclass{ws-rv9x6}

\usepackage[square]{ws-rv-van}             
\makeindex

\usepackage{epstopdf}

\begin{document}


\chapter{TASI Lectures on Astrophysical Aspects of Neutrinos}

\author[John F. Beacom]{John F. Beacom}

\address{Center for Cosmology and Astro-Particle Physics, \\
Departments of Physics and Astronomy, \\
The Ohio State University, \\
191 W. Woodruff Ave., Columbus, OH 43210, USA}

\begin{abstract}
  Neutrino astronomy is on the verge of discovering new sources, and
  this will lead to important advances in astrophysics, cosmology,
  particle physics, and nuclear physics.  This paper is meant for
  non-experts, so that they might better understand the basic issues in
  this field.
\end{abstract}

\body


\section{General Introduction}

It has long been appreciated that neutrino astronomy would have unique
advantages.  The principal one, due to the weak interactions of
neutrinos, is that they would be able to penetrate even great column
densities of matter.  This could be in dense sources themselves, like
stars, supernovae, or active galactic nuclei.  It could also be across
the universe itself.  Of course, the small interaction cross section
is also the curse of neutrino astronomy, and to date, only two
extraterrestrial sources have been observed: the Sun, and Supernova
1987A.  That's it.

However, a new generation of detectors is coming online, and their
capabilities are significantly better than anything built before.
Additionally, a great deal of theoretical effort, taking advantage of
the very rapid increases in the quality and quantity of astrophysical
data, has refined estimates of the predicted fluxes.  The basic message
is that the detector capabilities appear to have nearly met the
theoretical predictions, and that the next decade should see several
exciting first discoveries.

For these two talks, I was asked to introduce the topics of supernova
neutrinos and high-energy neutrinos.  See the other talks in this
volume for more about these and related topics.  To increase the
probability of this paper being read, I have condensed the material
covered in my computer presentation, focusing on the basic framework
instead of the details.  In the actual lectures, I made extensive use
of the blackboard, and of interaction with the students through
questions from them (and to them).  It isn't possible to represent
that here.  I thank the students for their active participation, and
hope that they've all solved the suggested problems!


\section{PART ONE: Supernova Neutrinos} 

\subsection{Preamble}

Over the centuries, supernovae, which appear as bright stars and then
disappear within a few months, have amazed and confused us.  We're
still amazed, and as Fermi said, we're still confused, just on a
higher level.  The historical observations of supernovae were of rare
objects in our own Milky Way Galaxy (here and elsewhere, ``Galaxy" is
used for the Milky Way, and ``galaxy" for the generic case).  Now that
we know their distances, we know that supernovae are extremely
luminous in the optical, in fact comparable to the starlight from the
whole host galaxy.  But that's not the half of it, literally.  If you
had neutrino-detecting eyes, you'd see the neutrino burst from a
single core-collapse supernova outshine the steady-state neutrino
emission from all the stars in a galaxy (the analog of solar
neutrinos) by a factor more like $10^{15}$ (that's a lot!).  This is
what enabled the detection of about 20 neutrinos from Supernova (SN)
1987A, despite its great distance.

A good general rule in decoding physical processes is ``Follow the
energy," much like ``Follow the money" for understanding certain human
endeavors.  For core-collapse supernovae, this means the neutrinos,
while for thermonuclear supernovae, this means the gamma rays.  These
are the direct messengers that reveal the details of the explosions.
In the following, I'll discuss this in more detail, mostly focusing on
the ``observational" perspective, since it's easy to be convinced that
observing these direct messengers is important, while hard to think of
how to actually do it.  As I will emphasize, this is much more than
just astronomy for its own sake: these data play a crucial role in
testing the properties of neutrinos, and more generally, in probing
light degrees of freedom beyond the Standard Model.


\subsection{Introduction}

Stars form from the collapse and fragmentation of gas clouds, and
empirically, the stellar Initial Mass Function is something like
$dn/dm \sim m^{-2.35}$, where $m = M_{\rm star} / M_{\rm sun}$, as
first pointed out by Salpeter in 1955, and refined by many authors
since.  You'll notice that this distribution is not renormalizable,
but don't start worrying about dimensional regularization -- a simple
cutoff near $m = 0.1$ is enough for our purposes.  What is the fate of
these stars?  There are two interesting broad categories.  The
``types" are observational distinctions, based on spectral lines, but
the divisions below are based on the physical mechanisms.

\begin{itemize}

\item {\bf Thermonuclear (Type Ia) supernovae} \\
  These have progenitors with $m \sim$ 3--8, and live for $\sim$ Gyr.
  The interesting case is when the progenitor has ended its nuclear
  fusion processes at the stage of being a carbon/oxygen white dwarf,
  while it has a binary companion that donates mass through accretion.
  Once the mass of the progenitor grows above the Chandrasekhar mass
  of $m = 1.4$, this carbon and oxygen will explosively burn all the
  way up to elements near iron, generating a tremendous amount of
  energy.  The most important isotope produced is $^{56}$Ni, which
  decays to $^{56}$Co with $\tau = 9$ days, which then decays to
  stable $^{56}$Fe with $\tau = 110$ days.  These decays produce MeV
  gamma rays and positrons that power the optical light curve.
  Indeed, a plot of luminosity versus time directly shows the two
  exponential components.

\medskip

\item {\bf Core-collapse (Type II/Ib/Ic) supernovae} \\
  These have progenitors with $m \sim$ 8--40, and live for less than
  $\sim$ 0.1 Gyr.  Importantly, the dynamics depend only on single
  stars, and not whether they happen to be in binaries or not.  As you
  know, the source of stellar energy is nuclear fusion reactions,
  which burn light elements into progressively heavier ones, until
  elements near iron are reached, and the reactions stop being
  exothermic.  Until that point, as each nuclear fuel is exhausted,
  the star contracts until the core is hot and dense enough to ignite
  the next one (remember, these reactions are suppressed by the Coulomb
  barrier).  The cutoff of $m \sim 8$ denotes the requirement of being
  able to burn all the way up to iron.  So what happens at that point?
  Once there is a $m \sim 1.4$ iron core, it is no longer generating
  nuclear energy, but it could support itself by electron degeneracy
  pressure, except for the fact that the massive envelope of the star
  is weighing down on it.  As discussed below, this leads to the
  collapse of the core and the formation of a hot and dense
  proto-neutron star, which cools primarily by neutrino emission over
  a timescale of seconds.

\end{itemize}

In both cases, a tremendous amount of energy is released in a time
that is very short compared to the lifetimes of stars, the resulting
optical displays are crudely similar, and shell remnants are left
behind.  For thermonuclear supernovae, the source of the energy is
nuclear fusion reactions, primarily revealed by the gamma rays from
nuclear decays.  The neutrino emission is subdominant, and no compact
remnant is left behind.  For core-collapse supernovae (often referred
to as type-II supernovae, even when this is inclusive of types Ib and
Ic as well), the source of the energy is gravitational, and is
primarily revealed by the neutrinos emitted from the newly-formed
neutron star (which may ultimately become a black hole).  There is
also gamma-ray emission, but it is subdominant compared to the
neutrinos.  Finally, one interesting fact is that for both categories
of supernova, the explosion energy is about $10^{51}$ erg, known as 1
``f.o.e." (fifty-one erg) or 1 ``Bethe."  Note that this is about
$10^{-3} M c^2$ for 1 solar mass of material.

The neutrino and gamma-ray emissions from supernovae could in
principle be detected from individual objects, or as diffuse glows
from all past supernovae.  Although low-mass stars are much more
common than high-mass stars, type Ia supernovae are more rare than
core-collapse supernovae by a factor of several, due to the
requirement of being in a suitable binary.  Before we get into
details, here's where things stand on observations of the direct
messengers.

\begin{itemize}

\item {\bf Gamma rays from thermonuclear supernovae} \\
  These have never been robustly detected from individual objects,
  though in a few cases the COMPTEL instrument set interesting limits.
  While a diffuse background of gamma rays is seen in the MeV range
  (and beyond), it is now thought that supernovae do not contribute
  significantly, making it more of a mystery what does.

\medskip

\item {\bf Neutrinos from core-collapse supernovae} \\
  These have been seen just once, from SN 1987A, but only with about
  20 events.  No diffuse background of neutrinos has been seen yet,
  placing interestingly tight limits on the contribution from
  supernovae.

\end{itemize}

For particle physicists, the primary interest is on two points.  If
neutrinos have unexpected properties, or if there are new light
particles that effectively carry away energy, then the neutrino
emission per supernova could be altered.  If there are processes in
the universe that produce MeV gamma rays, directly or after
redshifting, e.g., dark matter decay, then these may explain the
observed gamma-ray background.

Now let's turn to the basics of the neutrino emission from
core-collapse supernovae.  The gravitational binding energy release
can be simply estimated.  The gravitational self-energy of a
constant-density sphere is $(3/5) G_N M^2 / R$, and so
\begin{equation}
\Delta E \simeq \frac{3}{5} \frac{G_N M_{NS}^2}{R_{NS}}
- \frac{3}{5} \frac{G_N M_{NS}^2}{R_{core}}
\simeq 3 \times 10^{53} {\rm\ erg}
\simeq 2 \times 10^{59} {\rm\ MeV}\,,
\nonumber
\end{equation}
using the observed facts that neutron stars have masses of about $m =
1.4$ and radii of about 10 km.  Note that the second term in the
difference is negligible.  This is a tremendous amount of energy,
trapped inside a very dense object, and so no particles can escape and
carry away energy except neutrinos.  In fact, even the neutrinos must
diffuse out, as the density is high enough to counteract the smallness
of their interaction cross sections.

The core collapses until it reaches near-nuclear densities, at which
point it cannot proceed further, and hitting this wall creates an
outgoing shock.  If successful, the shock will propagate though the
envelope of the star, lifting it off and creating the optical
supernova.  If not, it will stall, and then the inflow of further
material will lead to black hole formation and no optical supernova.

The neutrinos are emitted from the core, within seconds of the
collapse, and carry nearly the full binding energy release noted
above.  It takes perhaps hours or days for the shock to break through
the envelope and begin the optical supernova, which is then bright for
months.  Importantly, the neutrinos are received {\it before} the
light.  It's not that they are tachyons, but rather just that they
were emitted first.  The kinetic energy of the supernova ejecta is
only $\sim 1\%$ of the total energy, and the energy in the optical
emission is even less.  The neutrinos are the most interesting, since
they carry most of the energy, are emitted in the shortest and
earliest time, and come from the densest regions.  Other than
gravitational waves, which have yet to be observed, only neutrinos can
reveal the inner dynamics of the core collapse process.

As noted, the neutrinos diffuse through the proto-neutron star,
meaning that they leave on a longer timescale and with lower energies
than they would if it were less dense.  It is typically assumed that
the neutrino emission per flavor (all six, counting neutrinos and
antineutrinos) is comparable.  That is, each takes about 1/6 of the
binding energy, and has thermal spectral with average energies of
10--20 MeV.  There is a vast literature about the differences between
flavors, and using this to test neutrino mixing, but this is beyond
our scope.

\begin{figure}[t]
\centerline{\includegraphics[width=3in,clip=true]{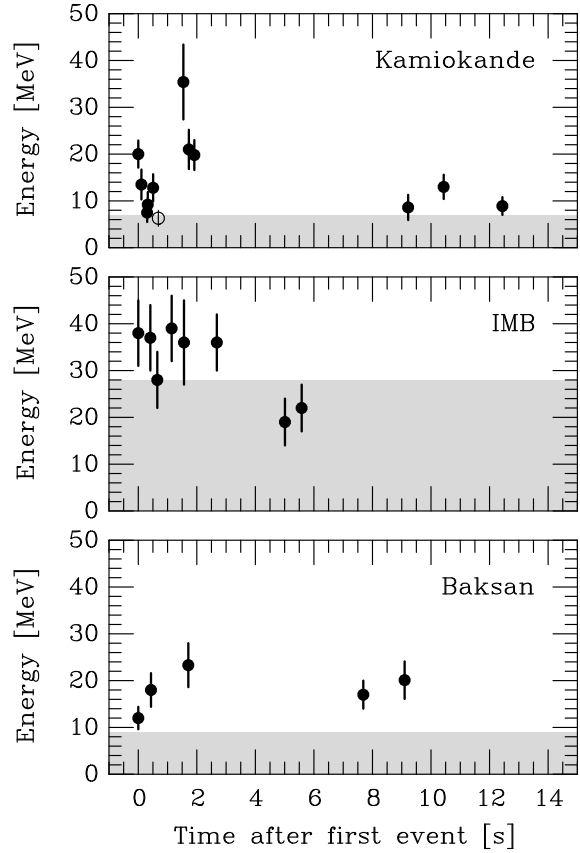}}
\caption{Scatterplots of the neutrino events associated with SN1987A,
  as seen in the Kamiokande~\cite{SN1987A-Kam},
  IMB~\cite{SN1987A-IMB}, and Baksan~\cite{SN1987A-Bak} detectors.
  The shaded regions indicate the nominal detector energy thresholds.
  Figure taken from Ref.~\cite{RaffeltReview}.}
\label{fig:SN1987A}
\end{figure}

The SN 1987A data are shown in Fig.~\ref{fig:SN1987A}.  These are
consistent with mostly being signal events due to inverse beta decay
on free protons, $\bar{\nu}_e + p \rightarrow e^+ + n$.  This reaction
channel is special due to its large cross section, and the fact that
the outgoing positron carries nearly the full antineutrino energy.
The other flavors are much harder to detect.  The first thing to
notice is that the duration of the burst was about 10 seconds.  The
second is that the typical energies were low tens of MeV.  (This is
complicated somewhat by the nontrivial response function of the
detectors, especially IMB, which was only effective at the highest
energies.)  At zeroth order, the Kamiokande and IMB data are
consistent with each other and theoretical expectations.  The Baksan
data are quite puzzling, as this detector was about ten times smaller
than Kamiokande, and thus they should have seen $\sim 1$ event;
probably detector backgrounds were present.

The most important message is that these data are consistent with the
picture of slow diffusion out of a very hot and dense object, i.e.,
with the birth of a neutron star, as is suggested also by the total
energetics, assuming a comparable neutrino emission per flavor.  You
can easily estimate the number of detected events yourself, using the
total energy noted above, the inverse beta cross section, and the
distance of 50 kpc.  (Interestingly, there is still no good
astronomical evidence for such a compact object in the SN 1987A
remnant.)  This kind of basic confirmation of the explosion mechanism
is what can do with such a small number of neutrino events.

How can we gather more supernova neutrinos?  There are three
possibilities.  First, {\bf Milky Way} objects, with $D \simeq 10$
kpc.  Taking into account the fact that we have much larger detectors
now, and assuming a typical distance in the Milky Way, we expect about
$10^4$ detected events in Super-Kamiokande.  Unfortunately, the
frequency is probably only 2 or 3 times per century, but we might get
lucky.  It will be very obvious if it happens.  Second, {\bf Nearby}
objects with $D \sim 10$ Mpc or less.  For these, one would need a
much larger detector, at the 1 Mton scale, and the number of detected
events per supenova is $\sim 1$.  On the other hand, the frequency is
about once per year.  To reduce backgrounds, this would require a
coincident detection of say two or more neutrinos, or one neutrino and
the optical signal.  Third, {\bf Distant} objects from redshifts $z
\sim$ 1--2 or less.  As a crude guide to how this works, imagine
supernovae at a distance such that the expected number of detected
events in Super-Kamiokande is $10^{-6}$.  Almost all of the time,
nothing happens, but for one supernova in a million, one neutrino will be
detected.  This seems crazy until you realize that the supernova rate
of the universe is a few per second.  Putting this together more
carefully leads to an expectation of several detected supernova events
per year in Super-Kamiokande (these will be uncorrelated with the
optical supernovae, due to the nearly isotropic nature of the
detection cross section).  A strong rejection of detector
backgrounds is required to make this work.

Of these three detection modes, I'll focus on the last, as it is the
least familiar.


\subsection{Supernovae in the Milky Way}

At present, the flagship supernova neutrino detector is
Super-Kamiokande, which is located in a deep mine in Japan.  It is the
largest detector with the ability to separate individual supernova
neutrino events from detector backgrounds.  Its huge fiducial volume
contains 22.5 kton of ultrapure water.  Relativistic charged particles
in a material emit optical \v{C}erenkov radiation, which is detected
by photomultiplier tubes around the periphery.

With $\sim 10^4$ events detected for a Milky Way supernova, the
Super-Kamiokande data could be used to map out the details of the
neutrino spectrum and luminosity profile.  Additionally, other
neutrino detection reactions, for which the yields are at the 1--10\%
level in comparison to inverse beta decay, would become important,
revealing more about the flavors besides $\bar{\nu}_e$.  The aspects
of detecting a Milky Way supernova are very interesting, and have been
extensively discussed elsewhere.


\subsection{Supernovae in Nearby Galaxies}

If Super-Kamiokande can detect $10^4$ events at a supernova distance
of 10 kpc, then it can expect to detect 1 event for a supernova
distance of 1 Mpc, somewhat larger than the distance to the M31
(Andromeda) and M33 (Triangulum) galaxies.  Unfortunately, a single
event isn't exciting by itself, and anyway, these galaxies appear to
have even lower supernova rates than the Milky Way.  Still, it makes one
wonder about greater distances.  The number of galaxies in each new
radial shell in distance increases like $D^2$, while the flux of each
falls like $1/D^2$.  As mentioned, one can beat even small Poisson
expectations with enough tries, so this is intriguing.

\begin{figure}[t]
\centerline{\includegraphics[width=3in,angle=0,clip=true]{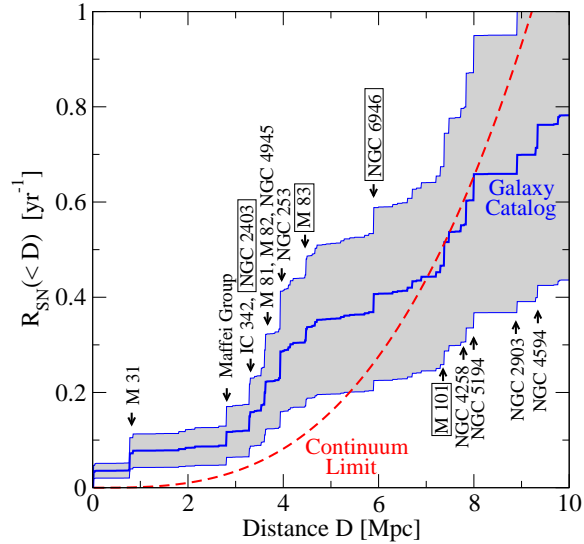}}
\caption{The predicted cumulative supernova rate for nearby galaxies
is shown by the blue line, and its uncertainty by the grey band
(together denoted as ``Galaxy Catalog").  The
redshift $z = 0$ limit of the cosmic supernova rate is also shown
(``Continuum Limit").  The observed local supernova rate in recent
years has been higher than either prediction.
Figure taken from Ref.~\cite{Peeping}.}
\label{fig:snrate}
\end{figure}

An estimate based on the known nearby galaxies shows that the
supernova rate with 10 Mpc should be about one per year, and this
is shown in Fig.~\ref{fig:snrate}.  In fact,
the observed rates in the past few years have been even higher.  A
detailed calculation shows that a larger detector than
Super-Kamiokande, something more on the 1 Mton scale, could detect
about one supernova neutrino per year.  (Such detectors are being
considered for proton decay studies and as targets for long-baseline
neutrino beams.)  That seems like a small rate, but bear in mind that
in the twenty years since SN 1987A, exactly zero supernova neutrinos
have been (identifiably) detected.  To reduce backgrounds, these
nearby supernovae would need a coincidence of at least two neutrinos or
one neutrino and the optical signal.  Perhaps most importantly, the
detection of even a single neutrino would fix the start time of the
collapse to about ten seconds, compared to the precision of about one
day that might be determined from the optical signal.  This would be
very useful for refining the window in which to look for a faint
gravitational wave signal.

Related to this is an effort called NO SWEAT (Neutrino-Oriented
Supernova Whole-Earth Telescope), led by Avishay Gal-Yam, to use a
network of telescopes worldwide to find all supernovae in nearby
galaxies.


\subsection{DSNB: First Good Limit}

The star formation rate was larger in the past, and in particular, was
about 10 times larger at redshift $z \simeq 1$ than it is today.
Since the lifetimes of massive stars are short, the core-collapse
supernova rate should closely follow the evolution of the star
formation rate, up to a constant factor.  This gives more weight to
distant supernovae than if the rate were constant.  On the other hand,
for supernova beyond $z \sim 1$, the neutrinos are so redshifted that
their detection probabilities are too low (at lower energy, the
detection cross section goes down while the detector background rates
increase).

Integrating the neutrino emission per supernova with the evolving
supernova rate, and taking into account the cosmological factors, the
accumulated spectrum of all past supernovae can be calculated.  This
is known as the Diffuse Supernova Neutrino Background (DSNB), or
sometimes as relic supernova neutrinos (which is a confusing and
deprecated term, i.e., these have nothing to do with the 2 K relic
background of neutrinos that decoupled just before big-bang
nucleosynthesis).

\begin{figure}[t]
\centerline{\includegraphics[width=3in,clip=true]{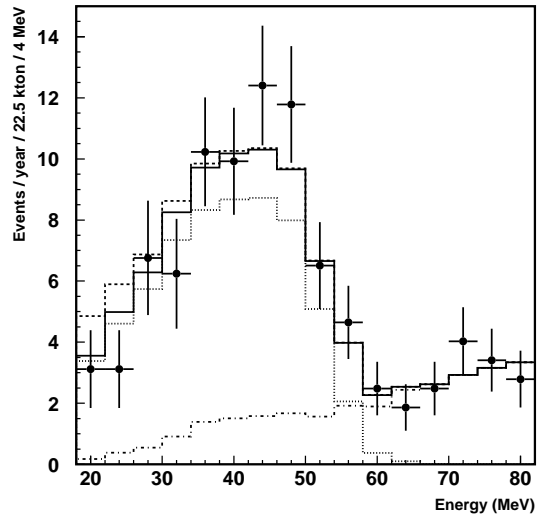}}
\caption{The event spectrum measured in Super-Kamiokande is
  denoted by the points with error bars.  The solid line indicates the
  expected total detector background rate (the dotted component is due
  to muon neutrinos, and the dot-dashed component is due to electron
  neutrinos).  The dashed line above the solid line indicates how
  large of an excess due to DSNB events could be present, given the
  statistical uncertainties.  Figure taken from Ref.~\cite{Malek}.}
\label{fig:SKrates}
\end{figure}

In 2000, a paper by Kaplinghat, Steigman, and Walker calculated the
largest plausible DSNB flux, and found it to be 2.2 cm$^{-2}$ s$^{-1}$
for electron antineutrinos above 19.3 MeV.  This was about 100 times
smaller than the existing limit from Kamiokande, so the prospects for
detection didn't look great.  Other calculations with reasonable
inputs (by modern standards) gave results that were a few to several
times smaller.

In 2003, the Super-Kamiokande collaboration published a limit that was
1.2 in the above units.  This was a milestone, because it showed for
the first time that there was hope of reaching the range in which a
detection might be made.  Still, as shown in Fig.~\ref{fig:SKrates},
there are large detector backgrounds that make it difficult to
identify the DSNB signal.  Note that for a background-limited search,
like this one, to improve the signal sensitivity by a factor of 3
takes a factor 9 more statistics.  Since this figure was based on 4
years of data, this would take a long time to collect (comparable to
the wait for a Galactic supernova!).


\subsection{DSNB: Detection with Gadolinium}

In order to make progress, it is necessary to find a way to eliminate
or at least severely reduce the detector background.  Mark Vagins (a
member of the Super-Kamiokande collaboration) and I decided to put our
heads together to find a way to isolate the DSNB signal.  This
resulted in a 2004 article in Physical Review Letters, though we were
forced to remove the code name of the project, ``GADZOOKS!," from the
title and text (but see the arXiv version).  Recall that the detection
reaction is $\bar{\nu}_e + p \rightarrow e^+ + n$, and that at
present, only the positron is detected.  We realized that the key was
to detect the neutron in time and space coincidence with the positron.
This is an old idea, and was used by Reines and Cowan in the first
detection of neutrinos (antineutrinos from a nuclear reactor).

\begin{figure}[t]
\centerline{\includegraphics[width=3in,clip=true]{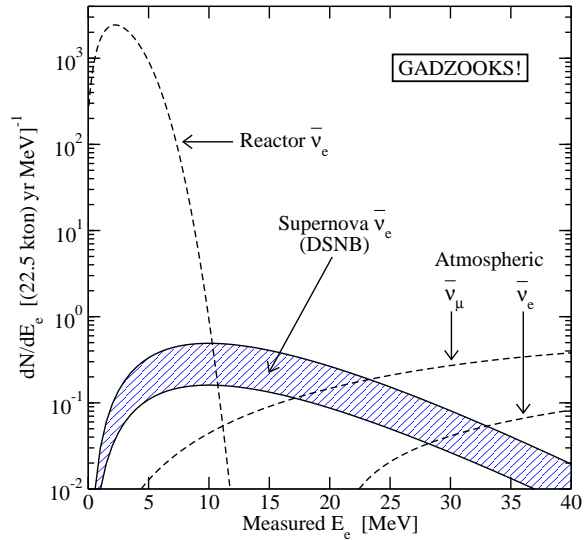}}
\caption{The DSNB signal and detector backgrounds expected
  in Super-Kamiokande if gadolinium is added.  Figure taken from
  Ref.~\cite{GADZOOKS!}.}
\label{fig:speclog}
\end{figure}

Saying that we had to detect the neutron was the easy part.  It was
more challenging to find a way to do this in a water-based detector,
where normally the neutrons capture on free protons.  That produces a
2.2 MeV gamma ray that Compton scatters electrons, but they are too
low in energy to be detectable.  We pointed out that the required
neutron tagging might be possible by using a 0.2\% admixture of
dissolved gadolinium trichloride (GdCl$_3$).  Gadolinium has a huge
neutron capture cross section, and produces an 8 MeV gamma-ray cascade
that reconstructs as an equivalent single electron of about 5 MeV,
which is readily detectable.

The really hard part was in establishing that this technique might be
possible in practice, which involved raising and answering many
difficult technical questions.  (Among them, finding a suitable
water-soluble compound of gadolinium.)  Somewhat to our surprise, we
found no obvious obstacles.  Mark Vagins has been leading a detailed
research and development effort, and so far, the prospects look very
good.

In Fig.~\ref{fig:speclog}, the spectra expected in Super-Kamiokande if
gadolinium is added are shown.  The atmospheric neutrino backgrounds
mentioned above are reduced by a factor of about 5.  Additionally,
backgrounds at lower energies are severely reduced, allowing the use
of a much lower threshold energy.  At moderate energies, it should be
possible to cleanly identify DSNB signal events.


\subsection{DSNB: Astrophysical Impact}

Now let's return to the predicted DSNB spectrum.  If either the
assumed star formation rate or the neutrino emission per supernova
were too large, then the predicted DSNB flux would already be ruled
out the the Super-Kamiokande data.

\begin{figure}[t]
\centerline{\includegraphics[width=3in,angle=-90,clip=true]{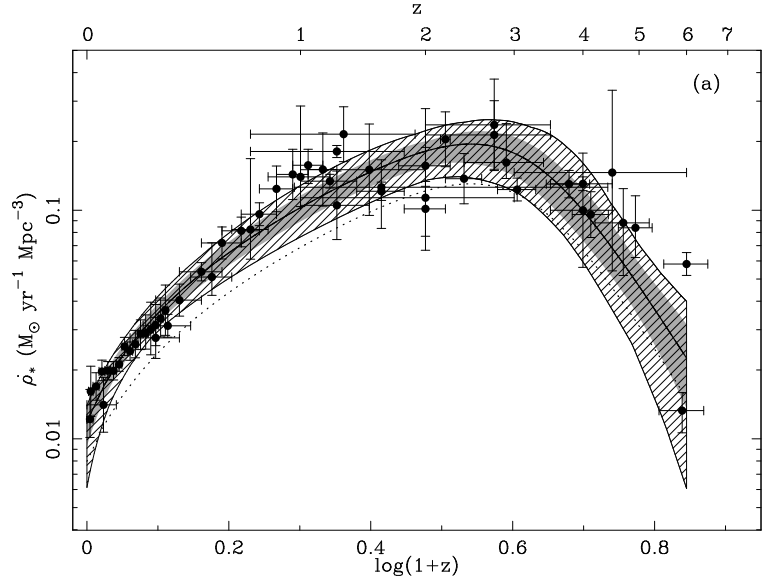}}
\caption{The star formation rate history, with selected data
  shown by points and the fit and uncertainty shown by the bands.
  Figure taken from Ref.~\cite{HopkinsBeacom}.}
\label{fig:SFR}
\end{figure}

Even since the time of the Super-Kamiokande limit, the astrophysical
data have improved substantially.  Andrew Hopkins and I synthesized a
wide variety of data to constrain the star formation and supernova
rate histories.  An example fit is shown in Fig.~\ref{fig:SFR}.  The
uncertainty band is much more narrow now than it was just a few years
ago.  The normalization of the cosmic star formation rate depends on
dust corrections.  If the true star formation rate were even somewhat
larger than determined here, then the DSNB neutrino flux would be too
large relative to the Super-Kamiokande limit.  The only way out would
be to require a substantially lower neutrino emission per supernova.

The corresponding calculated supernova rates are in good
agreement with data.   As an interesting aside, it was shown that
the diffuse gamma-ray background from type Ia supernovae is
too small to account for the observed data in the MeV range.  That
is particularly significant because many limits on exotic particle
physics depend on just this energy range.


\subsection{Back to the Scene of the Crime: SN 1987A}

If we now know the star formation history, then the only remaining
unknown is the neutrino emission per supernova.  Hasan Y\"uksel,
Shin'ichiro Ando, and I considered how well the Super-Kamiokande data
already restrict the neutrino emission per supernova.  The emission
models are usually parametrized in terms of the time-integrated
luminosity (or portion of the binding energy release) and the average
energy per neutrino (related to the temperature of the spectrum).  I
mentioned above that the Kamiokande and IMB data on the emission from
SN 1987A were mostly consistent.  In fact, when fitted with thermal
spectra, there are some discrepancies.

\begin{figure}[t]
\centerline{\includegraphics[width=3in,clip=true]{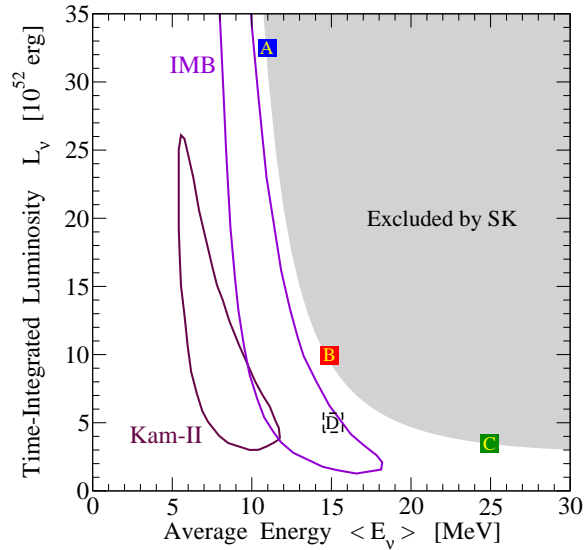}}
\caption{The contours labeled Kam-II and IMB are the allowed
  regions from the SN 1987A data, assuming a thermal
  spectrum~\cite{SN1987Afits}.  The shaded region is what is already
  excluded by the non-observation of a DSNB signal in
  Super-Kamiokande.  Figure taken from Ref.~\cite{Temperature}.}
\label{fig:temp-limit}
\end{figure}

In Fig.~\ref{fig:temp-limit}, I show that the DSNB data are probing
neutrino emission parameters only slightly larger than those deduced
from the SN 1987A data.  With reduced detector backgrounds, the DSNB
spectrum would be a new way to measure the neutrino emission per
supernova.

\subsection{Conclusions}

Why is understanding supernovae interesting and important?  For
particle physics, it is to test the properties of neutrinos, and to
search for new low-mass particles that cool the proto-neutron star.
For nuclear physics, it is to constrain the neutron star equation of
state and to shed light on the formation of the elements.  For
astrophysics, it is to understand the stellar life and death cycles
and the supernova mechanisms.  For cosmology, it is to
better understand the details of whether type Ia supernovae are
standard candles, and to probe the origins of the gamma-ray and
neutrino backgrounds.  With more data, we can't lose.


\newpage

\section{PART TWO: High-Energy Neutrinos}

\subsection{Preamble}

Now that we've covered the specific example of supernova neutrinos,
let's step back and comment on the general status and outlook in
neutrino astrophysics.

Unique among the Standard Model fermions, neutrinos are neutral, and
more generally, have only weak interactions.  This makes them
potentially sensitive to even very feeble postulated new interactions.
While the discovery of neutrino mass and mixing was ``new physics"
beyond the minimal Standard Model, the discovery of any new
interactions would be a much more radical step, as it would require
new particles as well.

This is one reason that we're interested in neutrinos.  The other,
already discussed, is that they will be especially powerful probes of
astrophysical objects, once these neutrinos are detected.  Already
with the neutrinos from the Sun and SN 1987A, the scientific return
was very rich: not only confirmation of the physics of their
interiors, but also a crucial piece in the discovery of neutrino mass
and mixing.  Ray Davis and Masatoshi Koshiba shared in the 2002 Nobel
Prize for this work, and their citation reads, ``...for pioneering
contributions to astrophysics, in particular for the detection of
cosmic neutrinos...."

The general achievements in neutrino physics in just the recent past
might be summarized as follows.  The {\bf cosmological} results are
the consistency of big-bang nucleosynthesis yields with three flavors
of neutrinos, and the exclusion of neutrinos as the (hot) dark matter.
In both cases, these facts have been established independently in the
laboratory as well.  The {\bf astrophysical} results are the discovery
of neutrinos from SN 1987A and the solution of the solar neutrino
problem.  The {\bf fundamental} results are the discovery of neutrino
mass and mixing, and the clear exclusion of a huge range of formerly
allowed models of exotic neutrino properties.

One of the lessons from this list is that we need data from new
sources to make new discoveries, and that those discoveries may have a
broader impact than initially thought.  Astrophysical sources reach
extremes of density, distance, and energy, and this will allow
unprecedented tests of neutrino properties, for example.

We can identify three frontiers where new sources will likely be
discovered soon.  By the rough energy scale of the neutrinos, we might
call these the MeV ($10^{-6}$ TeV) scale, the TeV scale, and the EeV
($10^{6}$ TeV) scale.  At the MeV scale, the focus is on the {\bf
  Visible Universe}, i.e., stars and supernovae, and Super-Kamiokande
is the main detector.  At the TeV scale, the focus is on the {\bf
  Nonthermal Universe}, i.e., jets powered by black holes, and the
primary detector is AMANDA, which is being succeeded by IceCube.  At
the EeV scale, the focus is on the {\bf Extreme Universe}, i.e., at
the energy frontier of the highest-energy cosmic rays, and one of the
key detectors is ANITA.


\subsection{Introduction}

Why do we think that high-energy neutrinos even exist?  First, because
cosmic rays (probably mostly protons) are observed at energies as high
as $10^{20}$ eV, and they are increasingly abundant down to at least
the GeV range.  Something is accelerating these cosmic rays, and it is
very likely that these sources also produce neutrinos.  Second,
because extragalactic gamma-ray sources have been observed with
energies up to about 10 TeV (and galactic sources up to about 100
TeV).  Again, something is producing these particles, and in large
fluxes, and it is likely that neutrinos are also produced.

So then why do we need neutrinos?  The problem with cosmic rays is
that they are easily deflected by magnetic fields, and so only their
isotropic flux has been observed, making the identification of their
sources very difficult.  The problem with photons is that they are
easily attenuated: a TeV gamma ray colliding with an eV starlight
photon is able to produce an electron-positron pair.  Thus at high
energies, only nearby objects can be seen.

High-energy neutrinos can be made through either proton-proton or
proton-photon collisions, depending on energies.  In either case,
pions are readily produced, and typically comparable numbers of
neutral and charged pions are made.  Neutral pions decay as $\pi^0
\rightarrow \gamma + \gamma$, and charged pions decay as $\pi^+
\rightarrow \mu^+ + \nu_\mu$, followed by $\mu^+ \rightarrow e^+ +
\nu_e + \bar{\nu}_\mu$ (with obvious changes for the charge
conjugate).  This is the {\bf hadronic} mechanism for producing gamma
rays and neutrinos.  There is also a {\bf leptonic} mechanism, based
on the inverse Compton scattering reaction $e^- + \gamma \rightarrow
\gamma + e^-$, where fast electrons collide with low-energy photons
and promote them to high-energy gamma rays.  Note that the leptonic
process produces no neutrinos.  It is a major mystery whether the
observed high energy gamma-ray sources are powered by the hadronic or
leptonic mechanism.  This is a key to uncovering the sources of the
cosmic rays.


\subsection{Sources and Detection at $\sim 1$ TeV}

\begin{figure}[t]
\centerline{\includegraphics[width=3in,clip=true]{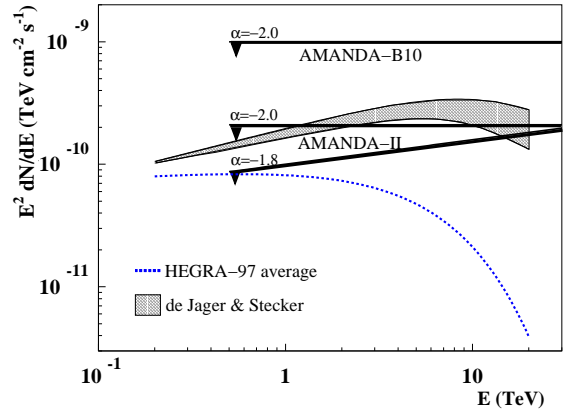}}
\caption{The dotted line is based on gamma-ray observations
  of the nearby AGN Markarian 501 by the HEGRA experiment, and the
  shaded band is a calculation that removes the assumed affects of
  attenuation en route. The labeled solid lines indicate AMANDA limits
  on the neutrino flux.  This object flares, and the gamma ray and
  neutrino data are not contemporaneous.  Figure taken from
  Ref.~\cite{Ahrens:2003pv}.}
\label{fig:mkn501}
\end{figure}

At the simplest level, hadronic sources produce nearly equal fluxes of
gamma rays and neutrinos (the corrections due to multiplicities, decay
energies, and neutrino mixing can be easily taken into account).
Therefore, the observed gamma-ray spectrum of an object like an AGN is
a strong predictor of the neutrino spectrum, if the source is hadronic
(if it is leptonic, then the neutrino flux will be zero).  Any
attenuation of the gamma-ray spectrum en route would mean that the
neutrino flux would be even larger.  An example is illustrated in
Fig.~\ref{fig:mkn501}, where it is shown that the neutrino detectors
are now approaching the required level of flux sensitivity.

For hadronic sources, the initial neutrino flavor ratios (adding
neutrinos and antineutrinos) are $\phi_e : \phi_\mu : \phi_\tau = 1 :
2 : 0$, following simply from the pion and muon decay chains.  After
vacuum neutrino mixing en route, these will become $\phi_e : \phi_\mu
: \phi_\tau = 1 : 1 : 1$.

Of all flavors, the muon neutrinos are the easiest to detect and
identify.  Through charged-current deep inelastic scattering
reactions, these produce muons that carry most of the neutrino energy,
and which have only a very small deflection from the neutrino
direction.  Muons and other charged particles produce optical
\v{C}erenkov radiation in the detector, which is registered by
photomultiplier tubes throughout the volume.  Muons produce
spectacular long tracks that can range through the kilometer of the
detector and beyond.  The detection of electron and tau neutrinos is
interesting and important too, but beyond our scope here.

To screen out enormous backgrounds from downgoing atmospheric muons,
these detectors only look for upgoing events.  Since muons cannot pass
through Earth, these muons must have been created just below the
detector by upgoing neutrinos.  Even after this, there are backgrounds
due to atmospheric neutrinos, themselves produced on the other side of
Earth, and thus hardly extraterrestrial.

\begin{figure}[t]
\centerline{\includegraphics[width=3in,clip=true]{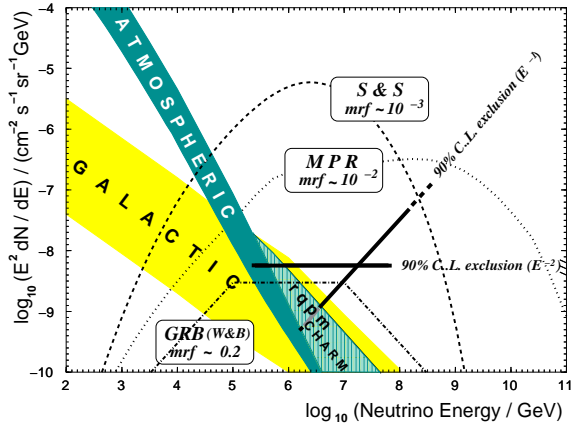}}
\caption{The sensitivity of IceCube is marked with heavy solid
  lines, as labeled.  The broken lines indicate various astrophysical
  diffuse flux models.  The shaded regions indicate the atmospheric
  neutrino, prompt/charm component thereof, and Galactic neutrino
  backgrounds.  Figure taken from Ref.~\cite{IceCube}.}
\label{fig:diffuse}
\end{figure}

An astrophysical point source can be identified as an excess in a
given direction, whereas the atmospheric neutrino background is
smoothly varying.  Transient point sources are even easier to
recognize.  On the other hand, diffuse astrophysical neutrino fluxes
are quite hard to separate from the atmospheric neutrino background.
The principal technique is that the former are believed to have
spectra close to $E^{-2}$, while the latter is closer to $E^{-3}$, and
steeper at higher energies.  Thus at high energies the astrophysical
diffuse fluxes should emerge as dominant.  Once cannot go too high in
energy -- the event rates get too low, and Earth becomes opaque to
neutrinos at around 100 TeV.  An example of the diffuse flux
sensitivity of IceCube is shown in Fig.~\ref{fig:diffuse}.

\subsection{Testing Neutrino Properties}

As an example of a novel neutrino property that could be tested once
astrophysical sources are observed, consider neutrino decay.  Why
should neutrinos decay?  Other than the fact that there is no known
interaction that can cause fast neutrino decay, why shouldn't they?
The other massive fermions all decay into the lowest-mass
generation in their family.  (Neutrinos can too, via the weak
interaction, but it is exceedingly slow.)  We'll consider simply
neutrino disappearance, i.e., that the other particle in the decay of
one neutrino mass eigenstate to another is too weakly interacting to
be detected.  It is quite hard to test for the effects of such decays.

Decay will deplete the original flux as
\begin{equation}
\exp\left(-t/\tau_{lab}\right) =
\exp\left(-\frac{L}{E} \times \frac{m}{\tau}\right)\,,
\nonumber
\end{equation}
where $L$ is distance, $E$ the energy, $m$ the mass, and $\tau$ the
proper lifetime.  For the Sun, the $\tau/m$ scale that can be probed
is up to about $10^{-4}$ s/eV.  On the other hand, for distant
astrophysical sources of TeV neutrinos, $L/E$ may be such that
$\tau/m$ up to about $10^{+4}$ s/eV is relevant!

How can we tell if decay has occurred, if the neutrino fluxes are
uncertain?  As mentioned, the flavor ratios after vacuum oscillations
are expected to be $\phi_e : \phi_\mu : \phi_\tau = 1 : 1 : 1$.
However, it is among the mass eigenstates, not the flavor eigenstates,
where decays take place.  Suppose that the heaviest two mass
eigenstates have decayed, leaving only the lightest mass eigenstate.
What is its flavor composition?  In the normal hierarchy, it has
flavor ratios $\phi_e : \phi_\mu : \phi_\tau \sim 5 : 1 : 1$, whereas
in the inverted hierarchy, they are $\sim 0 : 1 : 1$.  In either case,
they are quite distinct from the no-decay case, and the flavor
identification capabilities of IceCube should be able to distinguish
these possibilities.


\subsection{Sources and Detection at $\sim 10^6$ TeV}

Cosmic rays have been observed at energies above $10^{20}$ eV, and
there are no good answers as to what astrophysical accelerators may
have produced them.  However, this becomes even more puzzling when it
is noted that the universe should be opaque to protons above about $3
\times 10^{19}$ eV traveling over more than 100 Mpc.  There are no
obvious sources within that distance.

The process by which protons are attenuated is $p + \gamma \rightarrow
p + \pi^0, n + \pi^+$, where both final states are possible, and the
target photon is from the cosmic microwave background.  As with the
hadronic processes discussed above, the neutral pion decays produce
gamma rays and the charged pion decays produce neutrinos.  The gamma
rays are themselves attenuated, but the neutrino flux builds up when
integrating over sources everywhere in the universe.  Since the
attenuation process for the protons is called the GZK process
(Greisen-Zatsepin-Kuzmin), these are called GZK neutrinos.  Typical
energies are in the EeV range, and an isotropic diffuse flux is
expected.

\begin{figure}[t]
\centerline{\includegraphics[width=3in,clip=true]{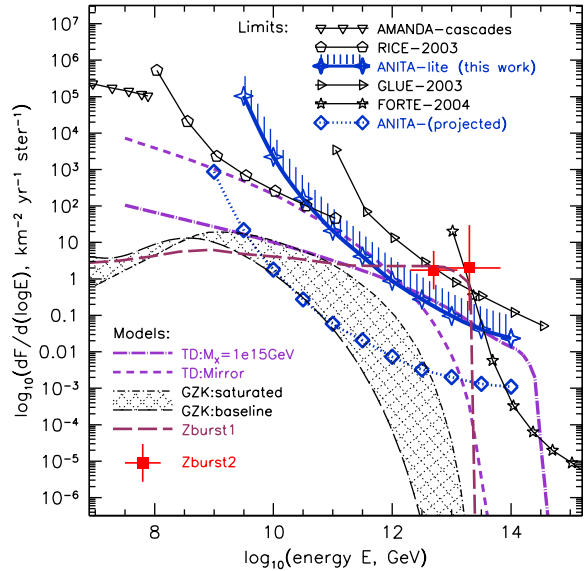}}
\caption{Real and projected neutrino flux sensitivities of various
  experiments (lines with points), along with various
  models (as labeled).  Figure taken from Ref.~\cite{ANITA}.}
\label{fig:ANITA}
\end{figure}

New experiments are being deployed to search for the GZK neutrino
flux, as shown in Fig.~\ref{fig:ANITA}.  Unlike IceCube, which is
based on optical \v{C}erenkov radiation, ANITA and other experiments
are based on radio \v{C}erenkov radiation that is emitted coherently
from the whole shower initiated by a neutrino in the ice or other
transparent medium.  ANITA is using the Antarctic ice cap as the
detector, and is observing it with radio antennas mounted on a
balloon.  So far, detector backgrounds appear to be negligible,
meaning that it should be straightforward to improve the signal
sensitivity with more exposure.

\begin{figure}[t]
\centerline{\includegraphics[width=3in,clip=true]{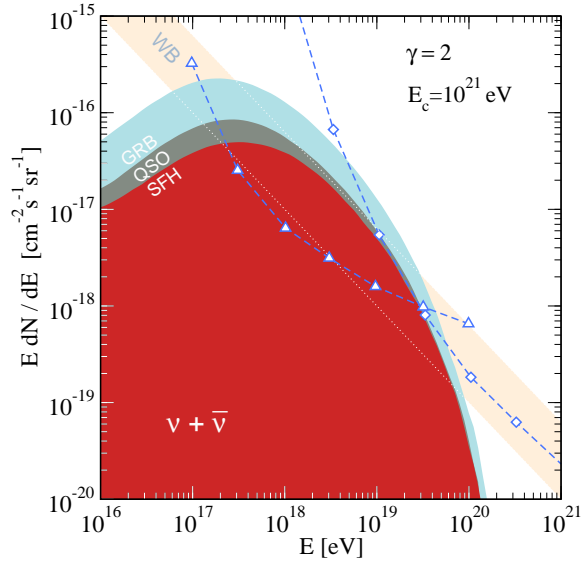}}
\caption{Predicted GZK neutrino fluxes, assuming that ultrahigh
  energy cosmic rays are produced in gamma-ray bursts, and according
  to how the latter rate evolves with redshift (i.e., following the
  star formation rate alone, or rising like that of the quasars, or
  depending on both the star formation rate and the evolving local
  metallicity).  The ``WB" band is the Waxman-Bahcall bound.  The
  curves with points are projected sensitivities for ANITA (upper) and
  ARIANNA (lower).  Figure taken from Ref.~\cite{YukselKistler}.}
\label{fig:grbgzk}
\end{figure}

In Fig.~\ref{fig:grbgzk}, I show the results of a very recent
calculation of the expected GZK neutrino fluxes.

Interestingly, when adjusted for the neutrino-quark center of mass
energy, the detection reactions are probing above the TeV scale,
opening the prospect of sensitivity to new physics in the detection
alone.

\subsection{Conclusions}

So far, zero high-energy astrophysical neutrinos have been detected.
However, the near-term prospects are very good, and are strongly
motivated by measured data on high-energy protons and photons.  Still,
this will not be easy, and large detectors with strong background
rejection will be needed.  If successful, these experiments will make
important astrophysical discoveries, e.g., whether gamma-ray sources
are based on the hadronic or leptonic mechanisms, the origins of
cosmic rays at all energies, etc.  We might even learn something new
about neutrinos in the process!


\section{Acknowledgments}

JFB was supported by National Science Foundation CAREER grant
PHY-0547102, and by CCAPP at the Ohio State University.



\end{document}